\documentclass[prc,twocolumn,showpacs,superscriptaddress]{revtex4}
\usepackage{graphicx,amsmath}
\def\om{\omega}

\def\lsim{\stackrel{\scriptstyle <}{\phantom{}_{\sim}}}
\def\gsim{\stackrel{\scriptstyle >}{\phantom{}_{\sim}}}
\def\rmd{{\rm d}}

\DeclareMathOperator{\rot}{rot}

\begin{document}

\title{Vorticity and hyperon polarization at NICA energies
}
\author{E. E. Kolomeitsev}
\affiliation{Matej Bel  University, SK-97401 Banska Bystrica, Slovakia}
\affiliation{Bogoliubov Laboratory for Theoretical Physics, Joint Institute for Nuclear Research, RU-141980 Dubna, Moscow region, Russia}
\author{V.D. Toneev}
\affiliation{Bogoliubov Laboratory for Theoretical Physics, Joint Institute for Nuclear Research, RU-141980 Dubna, Moscow region, Russia}
\author{V. Voronyuk}
\affiliation{Laboratory for Information Technologies, Joint Institute for Nuclear Research, RU-141980 Dubna, Moscow region, Russia}
\affiliation{Bogolyubov Institute for Theoretical Physics, Kiev, Ukraine}
\begin{abstract}
We study the formation of fluid vorticity and the hyperon polarization in heavy-ion collisions at NICA energies in the framework of the Parton-Hadron-String Dynamic Model, taking into account both hadronic and quark-gluonic (partonic) degrees of freedom.  The vorticity properties in peripheral Au+Au collisions at $\sqrt{s_{NN}}=$7.7\,GeV are demonstrated and confronted with other models. The obtained result for the $\Lambda$ polarization is in agreement with the experimental data by the STAR collaboration, whereas the model is not able to explain the observed high values of the anti-hyperon $\bar\Lambda$ polarization.
\end{abstract}
\maketitle

\section{Introduction}

The hydrodynamical approach has been applied to describe  heavy-ion reactions
for a long time~\cite{SCG82,CK86,SG86} and enjoys increasing interest in recent years. With a possibility of the quark-gluon plasma (QGP) formation in reactions at high beam energies, the scope of the hydrodynamical studies is widening even more~\cite{St05}. In peripheral heavy-ion collisions the initial angular
momentum  can be of the order of $(10^3-10^5)\hbar$, so that at the initial state of the hydrodynamical stage of the collision the shear flow pattern can be formed that could lead to rotation~\cite{CMS11} or even to the Kelvin-Helmholtz instability (KHI)~\cite{CSA12} in the reaction plane provided the medium has a low viscosity. Applications of modern computational schemes to these processes in (3+1) dimensions allow for a realistic description of the energy and momentum balance which leads to the observation of the collapse of a direct flow $v_1(y)$ and prediction of the third flow component or antiflow~\cite{CsR99}.

There is an inherent correlation between rotation of the medium and its magnetization~\cite{Heims-Jaynes62}, which may lead to particle-spin polarization. The primary example is the Einstein-de-Haas effect~\cite{EH15}, which demonstrates that sudden magnetization of electron spins in a ferromagnetic material leads to mechanical rotation because of the angular momentum conservation. The vorticity formation is largely discussed as a manifestation of the angular momentum conservation~\cite{BP07}. Barnett~\cite{Bar15} proved the existence of a reverse process - the
rotation of an uncharged body leads to the polarization of atoms and spontaneous magnetization. It is expected that quarks are also polarized in the rotating quark-gluon plasma (QGP) created in off-central heavy-ion collisions.
Liang and Wang first proposed that $\Lambda$ hyperons can be
polarized along the orbital angular momentum of two colliding nuclei \cite{LW06,GCD08}. Besides the global orbital angular momentum, the local vorticity may be created by a fast jet going through the QGP that will affect the hadron polarization as well~\cite{BGT07}.

The method of computing  spin polarization in the matter near equilibrium was developed within statistical hydrodynamics approach in Refs.~\cite{BecPR08,BecP08,Bec13}. It was later  confirmed within the quantum-kinetic approach~\cite{FPWW16}. Relativistic fluid dynamics of a particle with spin was also reconsidered recently in Ref.~\cite{FFJS17}.

Some hydrodynamic calculations quantitatively predict the global polarization in off-central heavy-ion collisions~\cite{BecCW13,KBec17,XBSWC16,XWC17,CMW13,KB17}.
The fluid vorticity creatred in heavy-ion collisions has also been investigated in transport simulations \cite{CWBS14,Bec15,SK17}. For more studies
of the fluid vorticity and $\Lambda$ polarization we refer the reader to
Refs.~\cite{PPWW16,AFGK16,BGST13,TU15,IS17} and the review article~\cite{QW17}.

Recently, STAR measured the global polarization of $\Lambda$ and $\bar\Lambda$ in off-central Au+Au collisions in the Beam Energy Scan (BES) program~\cite{BES}. From the measured
polarization, the fluid vorticity of the strongly coupled QGP and the magnitude of the magnetic field created in off-central heavy-ion collisions were extracted for the first time using the spin-vorticity and spin-magnetic coupling~\cite{BES}. It indicates that the rotational fluid has the largest vorticity that ever existed in the universe of the order of 
$10^{-21}\,{\rm Hz}$. So the strongly coupled QGP has an additional extreme feature: it is the fluid with the highest vorticity.
The observation of polarization of hyperons plays an important role in probing the vorticity field of the QGP. Therefore, it is worth studying the inherent correlation between the hyperon polarization and the microscopic vortical structure in detail.

The vorticity developed in high-energy heavy-ion collisions was estimated within various models. Recently, a comprehensive study of the $\Lambda$ polarization of the RHIC beam-energy scan was presented in~\cite{KBec17} where only the lowest RHIC energy overlaping with the NICA energy range was considered. In this paper, we focus on the properties of the vorticity field and the global $\Lambda$ polarization at the energy $\sqrt{s}=7.7$ GeV within the
Parton-Hadron-String Dynamics (PHSD) model which is proved to work reliably at the NICA energies~\cite{HSD,PHSD}. The PHSD transport approach~\cite{PHSD,LBC16} is a microscopic covariant dynamical model for strongly interacting systems formulated on the basis of the Kadanoff-Baym equations for Green’s functions in the phase-space representation. The approach consistently describes the full evolution of  relativistic heavy-ion collision from the initial hard scatterings and string formation through the dynamical deconfinement phase transition to strongly-interacting quark-partonic quasiparticles as well as hadronization and the subsequent interactions in the expanding hadronic phase as in the Hadron-String-Dynamics transport approach~\cite{HSD}.

The used PHSD version is extended to incorporate essential aspects of chiral symmetry restoration (CSR) in the hadronic sector (via the Schwinger mechanism)~\cite{PCSSMB16}. The calculated data are discussed and compared with the results of other models for the same energy range.

\section{Definitions}

To compute vorticity, we must first define numerically the realistic velocity field for nuclear collisions.
The kinetic model tracks positions and momenta of all particles at any moment of time. These particles need to be fluidized on space-time grids in order to calculate the velocity field numerically. This can be achieved by introducing a grid in the coordinate space and a smearing function $\Phi(\vec{x}, \vec{x}_i(t))$ for each particle where $\vec{x}$ is the field point and $\vec{x}_i(t)$ is the time-dependent coordinate of the $i$th particle. The effect of $\Phi(\vec{x}, \vec{x}_i(t))$ is to smear a physical quantity, e.g., energy or momentum, carried by the $i$th particle what located at $\vec{x}_i(t)$, to another coordinate point $\vec{x}$. Therefore, $\Phi(\vec{x}, \vec{x}_i)$ somehow represents the quantum wave packet of the $i$th particle. So the particle distribution function can be written as
\begin{align}
f(t,\vec{x}, \vec{p}) =\sum_i  (2\pi)^3 \delta^{(3)}(\vec{p} - \vec{p_i}(t))\frac{1}{N}
\Phi(\vec{x}, \vec{x}_i(t))
\end{align}
where $N =\int d^3x \Phi(\vec{x}, \vec{x}_i(t))$ is a normalization factor,
$p_i(t)$ and $p^0_i(t)$ are the momentum and energy of the $i$th particle, and the summation
is over all the particles on the grid.
Then the smeared (averaged) energy-momentum
tensor and particle number current are given by
\begin{align}
\label{Tmunu}
T^{\mu\nu}(t,\vec{x}) &= \int \frac{d^3p}{(2\pi)^3} \frac{p^\mu p^\nu}{p^0 }f(t,\vec{x}, \vec{p})
\nonumber\\
 &= \frac{1}{N} \sum_i \frac{p_i^\mu(t) p_i^\nu(t)}{p_i^0(t) } \Phi(\vec{x}, \vec{x}_i(t)) ,  \\
J^\mu(t,\vec{x}) &=  \int \frac{d^3p}{(2\pi)^3} \frac{p^\mu }{p^0 }f(t,\vec{x}, \vec{p})
\nonumber\\
&=
 \frac{1}{N} \sum_i \frac{p^\mu_i(t) }{p_i^0(t) } \Phi(\vec{x}, \vec{x}_i(t)) ,
\label{curr}
\end{align}
In our simulations,
$p_i(t)$ and $x_i(t)$ in each event and at each time moment are generated by the PHSD model~\cite{PHSD}.

The collective velocity field associated with the particle flow (\ref{curr}) in the given cell $a$ is defined as
\begin{align}
v^a(t,\vec{x}) =\frac{J^a(t,\vec{x})}{J^0(t,\vec{x})}.
\end{align}
where $a$ = 1, 2, 3 are the spatial indices. So, the velocity field for a given colliding event on some grid is \cite{DH16},
\begin{align}
v^a (t,\vec{x}) = \frac{1}{\sum_i\Phi(\vec{x}, \vec{x}_i(t))} \sum_i
\frac{p^a_i(t)} {p^0_i(t)} \Phi(\vec{x}, \vec{x}_i(t)),
\end{align}

Unlike classical hydrodynamics, where vorticity is defined as $\rot \vec{v}$ only, in relativistic hydrodynamics, one can introduce several different vorticities, each useful in different applications. We will use the definitions from~\cite{DH16}.In the Eckart frame \emph{the kinetic vorticity} is defined
as
\begin{align}
\omega_{\mu\nu}=\frac12 (\partial_\nu u_\mu -\partial_\mu u_\nu)\,,
\end{align}
where $u_\nu$ is a relativistic four-vector of the velocity field
\begin{align}
u_\nu(t,\vec{x})=\gamma(1,\vec{v}(t,\vec{x}))\,,\quad \gamma(t,\vec{x})=\frac{1}{\sqrt{1-\vec{v}\,^2(t,\vec{x})}}
\end{align}
in the fluid rest frame. The spatial components of the kinetic vorticity can be written in terms of the circulation of velocity
$\om_{ij}=\frac12\epsilon_{ijk}(\rot\vec{v})_k$ and the mixed components are
$\om_{0j}=\frac12(\vec{\om}_0)_j$, where we introduce a vector
\begin{align}
\vec{\om}_0= \frac12(\vec{\nabla} u_0+\partial_t \vec{u})\,.
\end{align}

We shall use also \emph{the thermal vorticity} which
is defined as
\begin{align}
\label{th-vel}
\beta_\mu=\frac{u_\mu}{T},
\end{align}
i.e, the field of the reciprocal temperature flow
\begin{align}
\label{Tst}
\varpi_{\mu\nu}=\frac12 (\partial_\nu \beta_\mu -\partial_\mu \beta_\nu)\,.
\end{align}
Thermal vorticity $\varpi_{\mu\nu}$ is dimensionless and, in contrast to relativistic  vorticity, depends on the temperature gradients and,
as shown in~\cite{Bec13}, determines the induced polarization vector
of relativistic particles with spin. As for the case of the kinetic
vorticity we can introduce the circulation of the vector $\vec{\beta}$, $\vec{\varpi}_{ij}=\frac12\epsilon_{ijk}\rot\vec{\beta}$ and the vector
\begin{align}
\vec{\varpi}_0= \vec{\nabla} \beta_0+\partial_t \vec{\beta}
\end{align}
determining $\varpi_{0j}=\frac12(\vec{\varpi}_0)_j$\,.

\section{Numerical results for vorticity formation}

\begin{figure*}
\includegraphics[width=15cm]{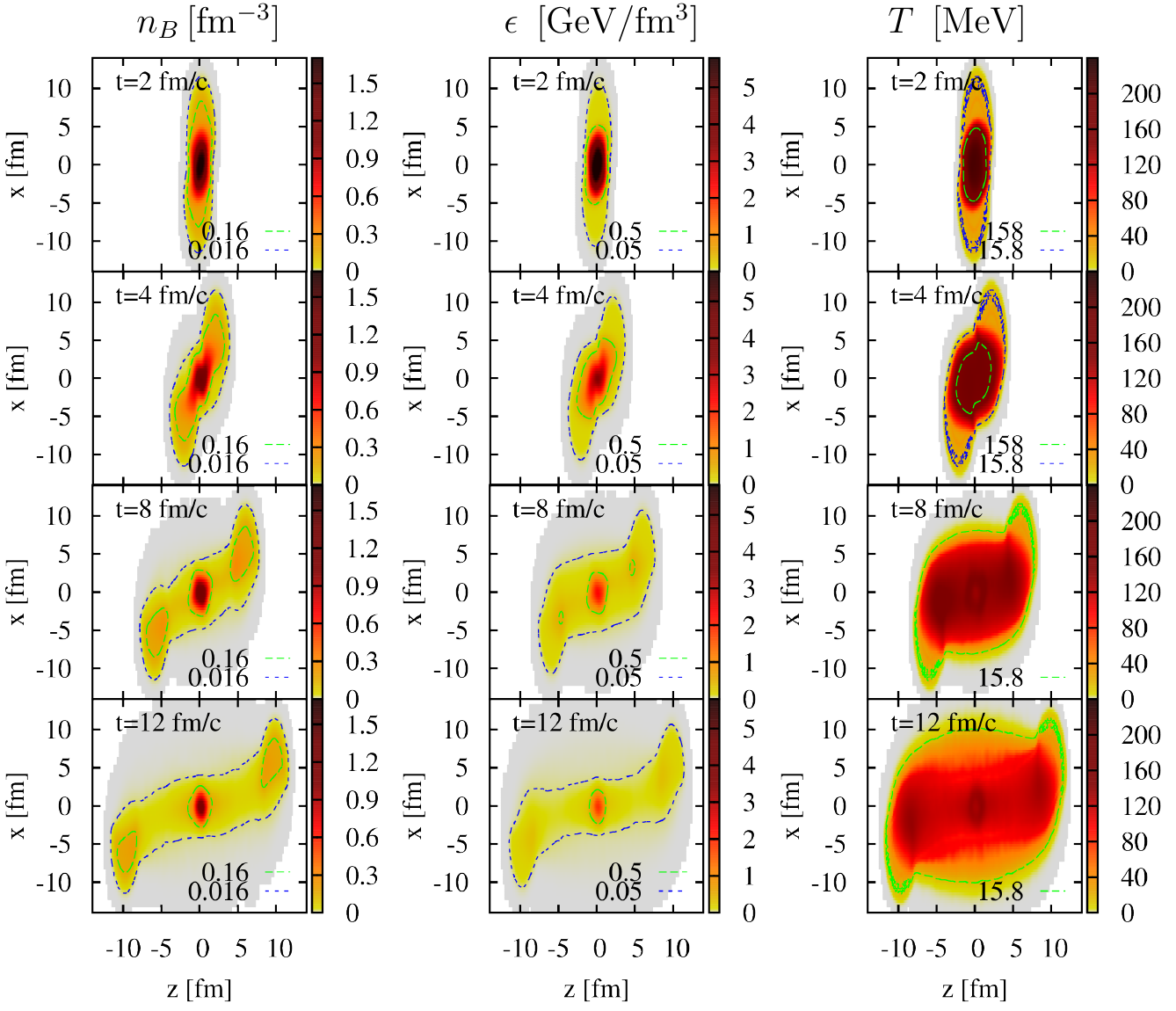}
\caption{(Color-online) Time dependence of the baryon density, energy density and temperature in Au$+$Au ($\sqrt{s}=7.7$\,GeV) collisions with the impact parameter 7.5\,fm (local system).
 }
\label{BarT}
\end{figure*}

The evolution of quantities characterizing a state of nuclear matter formed in Au+Au collision -- baryon density $n_B$, energy density $\epsilon$ and temperature $T$ -- is shown in Fig.~\ref{BarT}. The time evolution is calculated within the PHSD model ~\cite{PHSD} without introducing a freeze-out procedure. The first time-moment $t=2\,{\rm fm}/c$ corresponds to the case when the centers of colliding nuclei approaching each other at an almost minimal distance, then a fluid is formed and fluid matter expands developing a less dense shell in the peripheral zone. The freeze-out takes some finite time and occurs locally in cells specified by freeze-out energy density $\sim 0.2\,{\rm GeV/fm}^3$~\cite{Iv13}. As is seen, in our model such a regime is reached somewhere between $t=$8 and 12\,${\rm fm}/c$. The hybrid model joining the  kinetic and hydrodynamic descriptions gives for this case about 10\,fm/c for an average formation time~\cite{KhT07}.

The temperature $T$ entering into Eq.~(\ref{Tst}) is not defined within the kinetic PHSD approach. To find it, we use the same grid as for a vorticity study and in every cell solve the equations for conservation of the energy
and baryon charge. The evolution of the average energy, temperature and baryon density is shown in Fig.\ref{BarT}. All these distributions are rather smooth, because  the relativistic EOS for a mixture of the ideal resonance-gas (with nuclear potential  for baryons) and partons is used ~\cite{KhT07}.

\begin{figure*}
\centering
\includegraphics[width=15cm]{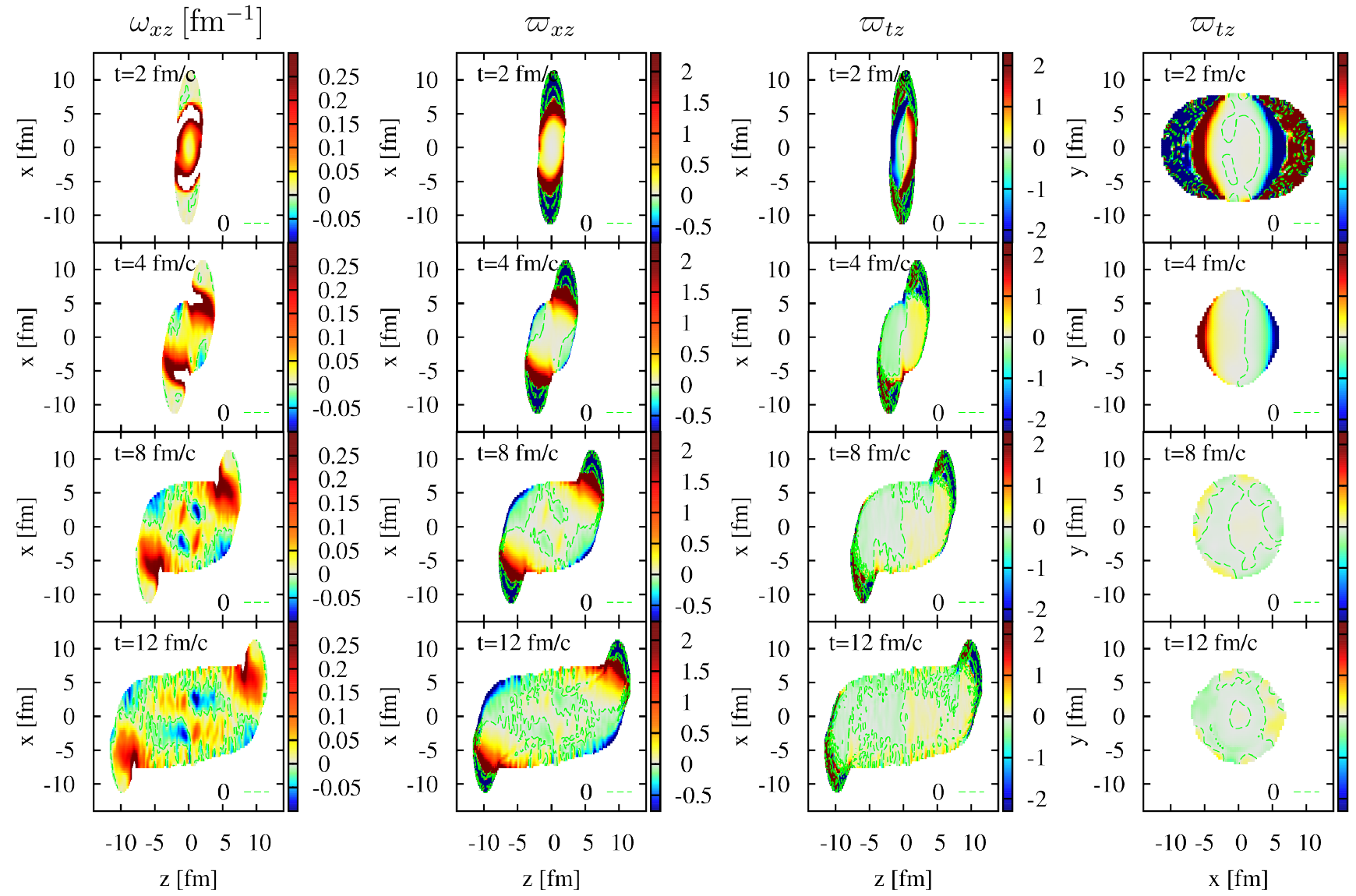}
\caption{(Color-online) Time dependence of the relativistic vorticity $\omega$ (the first column) and different projections of the thermal vorticity $\varpi_{xz}$ in Au+Au ($\sqrt{s}=7.7$\,GeV) collisions with the impact parameter 7.5\,fm. Thin contour lines correspond to the boundary $\varpi_{xz}=$0.  }
\label{Vortt}
\end{figure*}

The evolution of the relativistic kinematic vorticity in the reaction plane, $\omega_{xz}$, and different components of relativistic thermal vorticity, $\varpi_{xz}$, calculated in the PHSD model for 8400 events is presented in Fig.~\ref{Vortt}, cf.~\cite{IS17,KBec17}. The results are obtained under the condition $\epsilon >0.1 \epsilon_0= 0.015\,{\rm GeV/fm}^3$.
As seen in column 2 and 3 of Fig.~\ref{Vortt}) the thermal vorticity in the $(x,z)$ plane, $\varpi_{xz}$, is larger on the boundary of the system compared with the relativistic vorticity (column 1 in Fig.~\ref{Vortt})
because of the smallness of $T$ and large gradients. In peripheral collisions particle multiplicities are relatively small; therefore, fluctuations
in the reaction plane are considerable. The relativistic $\omega$ and thermal $\varpi$ vorticities fluctuate strongly at the final stage of interaction. These random fluctuations are visible at later times in the dilute matter and, especially, on outer edges of the fireball where the thermal vorticity has an enhanced amplitude. The $(x,y)$ projection of $\varpi_{tz}$ demonstrates clear cylindric symmetry (see  column 4 in Fig.~\ref{Vortt}).

\begin{figure*}
\centering
\includegraphics[width=13.5cm]{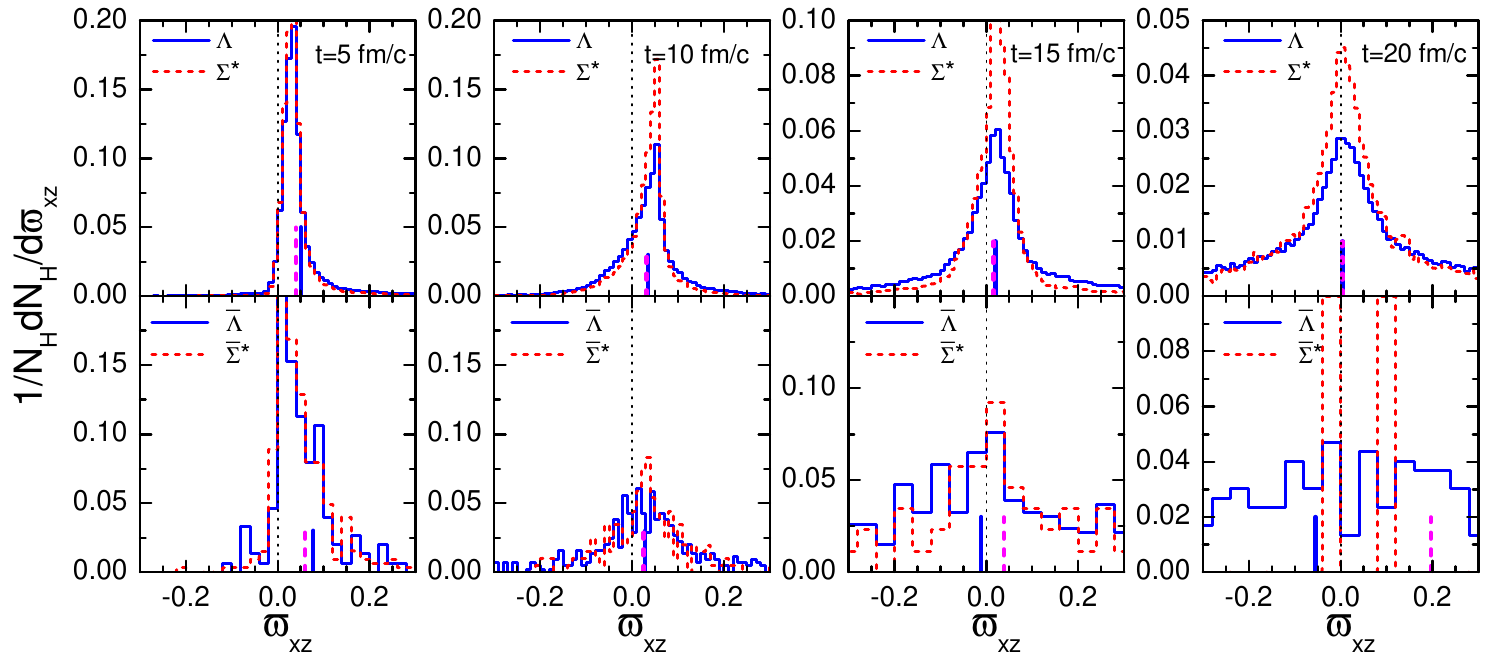}\\
\caption{(Color-online) Thermal vorticity distribution of $\Lambda$ and  $\Sigma^*$ hyperons (upper row) and $\overline{\Lambda}$ and $\overline{\Sigma^*}$ anti-hyperons (lower row) at 4 interaction time moments in Au+Au ($\sqrt{s}=7.7$\,GeV) collisions with the impact parameter $b=7.5$\,fm.
The vertical lines indicate the averaged values of the thermal vorticity.}
\label{omT-dist}
\end{figure*}

\begin{figure*}
\centering
\parbox{6.6cm}{\includegraphics[width=6.6cm,clip]{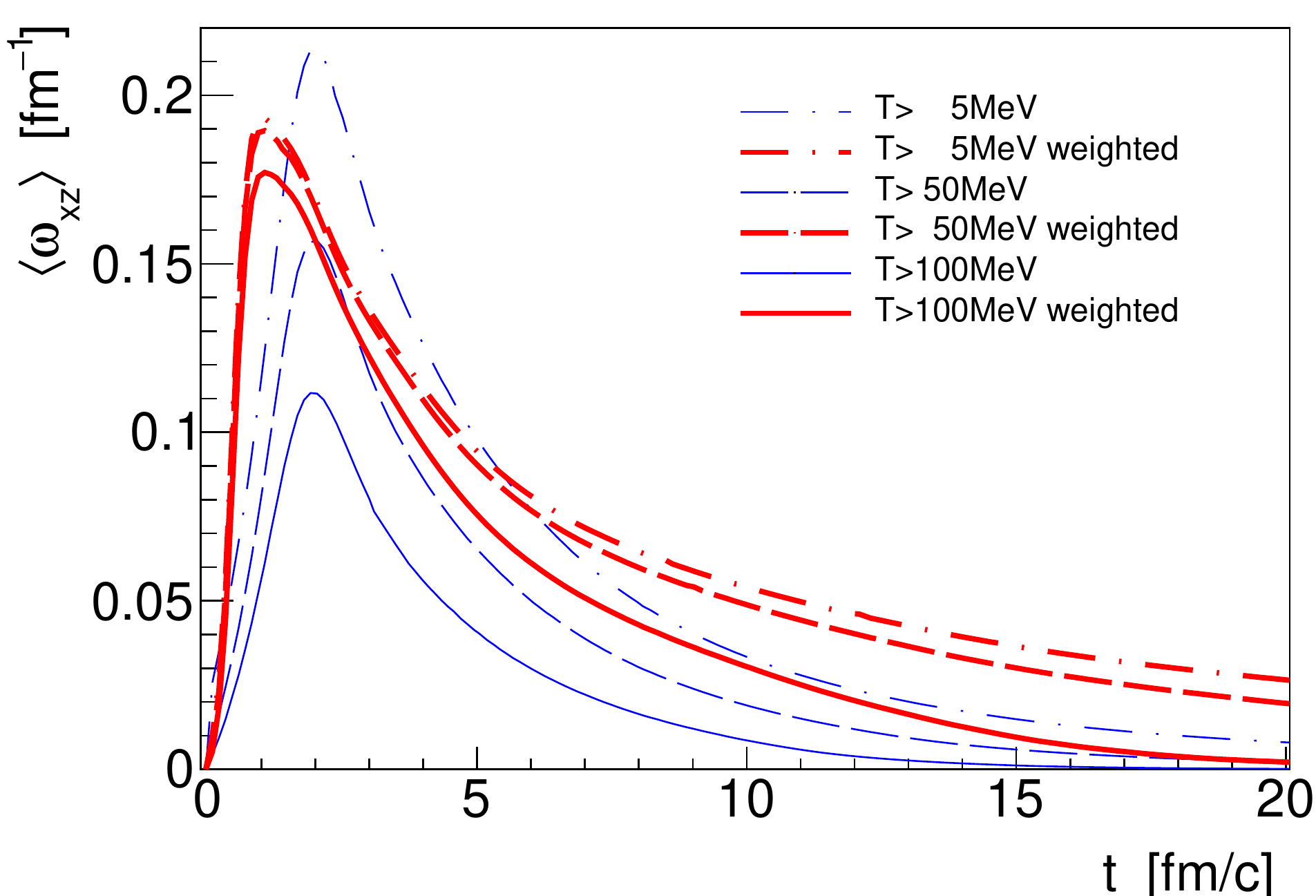}}
\parbox{6.5cm}{\includegraphics[width=6.5cm,clip]{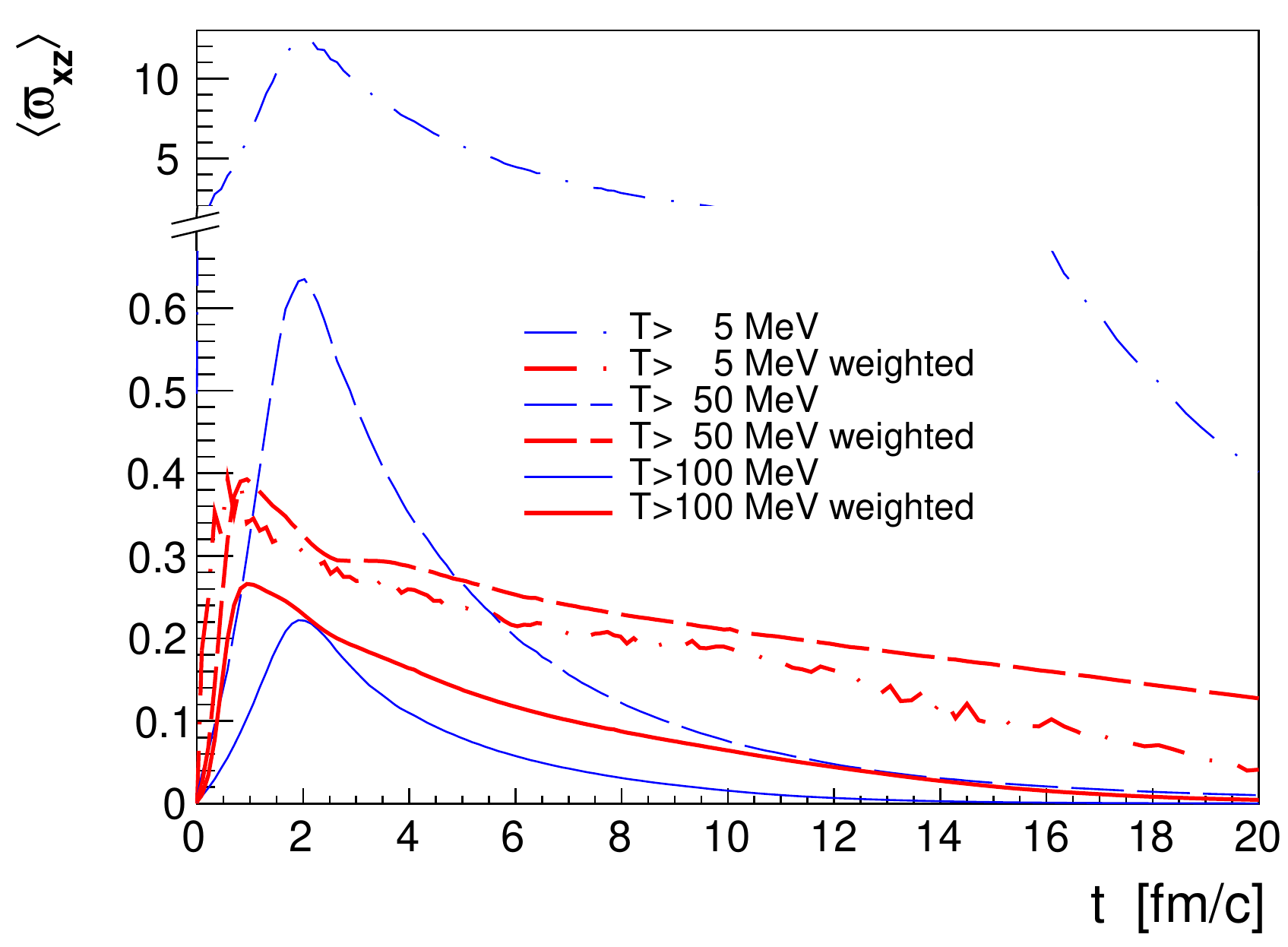}}
\caption{(Color-online) Time dependence of weighted and non-weighted kinetic $\left<\omega_{xz}\right>$ (left panel) and thermal $\left<\varpi_{xz}\right>$ (right panel)  vorticities of strange hadrons. Thin lines are plotted for non-weighted quantities, solid lines -- for weighted ones.}
\label{weighted-vor}
\end{figure*}

\begin{figure*}
\centering
\includegraphics[width=14.5cm,clip]{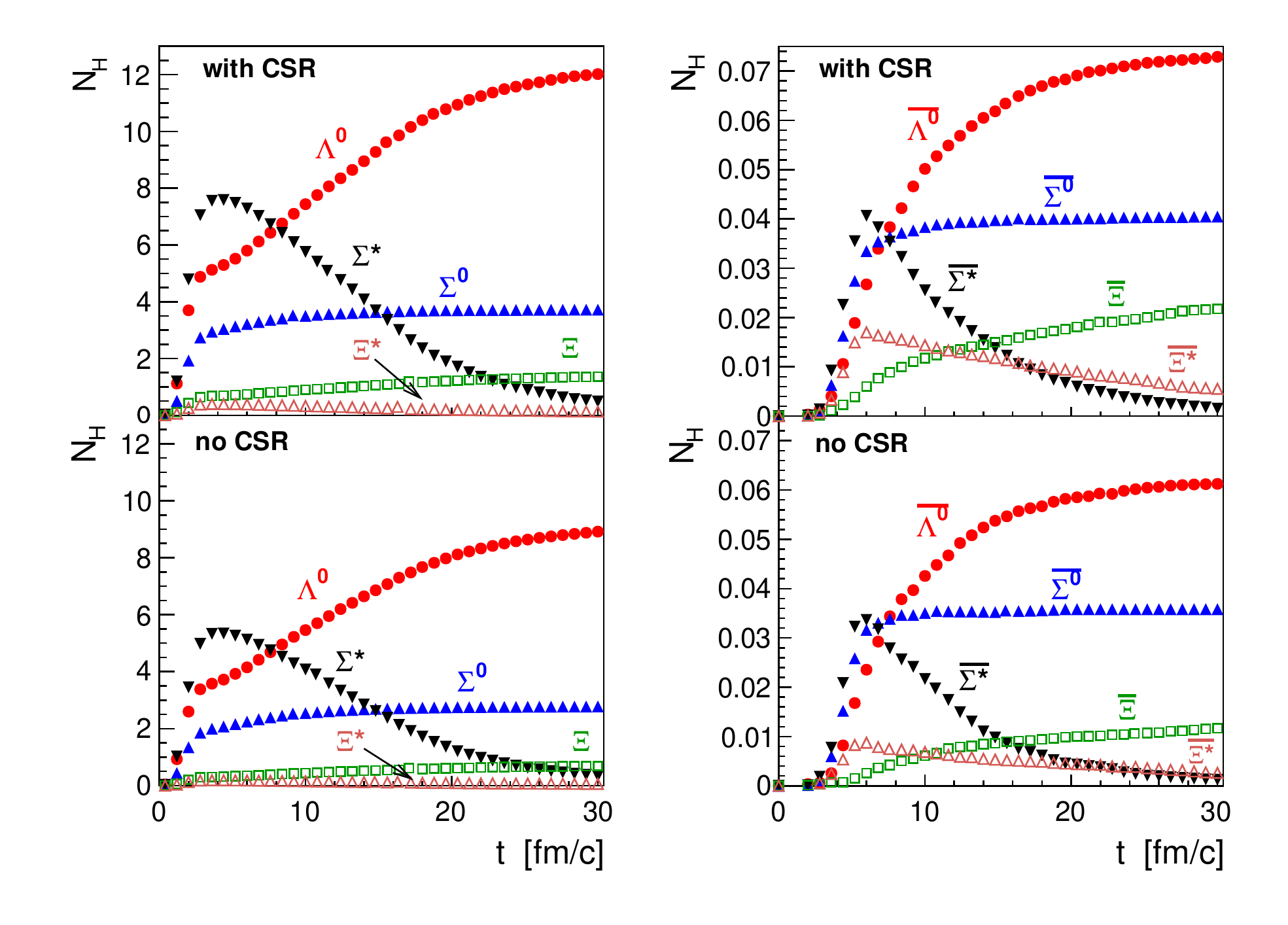}
\caption{(Color-online) Average numbers of strange particles (left panel) and anti-particles (right panel) as functions of collision time for Au+Au ($\sqrt{s}=7.7$\,GeV) with the impact parameter $b=7.5$\,fm. The upper panels correspond to the results calculated with and the bottom ones without the chiral symmetry restoration effect~\cite{PCSSMB16}. }
\label{Lambda-num}
\end{figure*}

In order to illustrate which vorticity field is seen by hyperons, we plot in
Fig.~\ref{omT-dist} the thermal vorticity distributions of $\Lambda$ and $\Sigma^*(1385)$ hyperons and their anti-particles for various moments of time. The histograms are normalized to unity in the interval of vorticities $|\varpi_{xz}|\le 1$. We see that the distributions are typically asymmetric with respect to zero with a positive averaged value. The averaged thermal vorticity for $\Lambda$ is typically larger than that for  $\Sigma^*$. At initial times $t\sim 5\,{\rm fm}/c$ the averaged vorticity for $\Lambda$ is maximal and then decreases monotonously, whereas the averaged vorticity for $\Sigma^*$ increases first up to $t\sim 10\,{\rm fm}/c$ and decreases at later times.The distributions for $\Sigma^*$ are typically broader than those for $\Lambda$. For the anti-hyperons the general pattern is similar for times $t\lsim 10\,{\rm fm}/c$. For later times the number of anti-hyperons dcreases rapidly and the distributions show strong fluctuations. The averaged vorticity values for $\overline{\Lambda}$ and $\overline{\Sigma^*}$ can become negative for $t\gsim 15\,{\rm fm}/c$.

To reduce fluctuations due to the regions, where the matter density is quite low and the hydrodynamic description is less applicable, one considers a proper-energy-density-weighted kinematic and thermal vorticities in the whole volume and in the reaction $(xz)$ plane.
\begin{align}
\label{Eaver}
\langle\varpi_{xz}(\vec{x})\rangle = \frac{\sum_\alpha \varpi_{xz}^\alpha \epsilon_{xz}^\alpha}
{\sum_\alpha \epsilon_{xz}^\alpha}.
\end{align}
weighted with the local energy density in every $\alpha$ cell at the given fixed vorticity. The cells with the values of $\langle\varpi_{xz}\rangle$ smaller than some threshold value are rejected in the sum. Such an averaging procedure is used in Ref.~\cite{IS17} but it differs from that in Refs.~\cite{CMW13,CWBS14,TU15}.

Keeping in mind that $\Lambda$ hyperons are abundantly produced from the hottest region of the system, it is of interest to apply certain constrains on this averaging. Let us consider the weighting of cells with $T>T_m$ for different values of the cut-temperature $T_m$ as
\begin{align}
\label{aver-con}
\langle\varpi_{\mu\nu}(t)\rangle_{T>T_m}=\frac{\int_{T>T_m} d^3x \varpi_{\mu\nu}(t,\vec{x})\, \epsilon(t,\vec{x})}{ \int_{T>T_m} d^3x \, \epsilon(t,\vec{x})}.
\end{align}
Time evolution of the quantity $\langle\varpi_{\mu\nu}(t)\rangle_{T>T_m}$ is presented in Fig.~\ref{weighted-vor} for $T_m=$5, 50 and 100 MeV for both
kinetic and thermal averaged vorticities. The inclusion of the weighting procedure with a threshold temperature $T_m$ strongly suppresses peaks of the thermal vorticity at early times, especially, for small $T_m$, but this effect is noticeably weaker for the weighted kinetic vorticity (see left panel in Fig.~\ref{weighted-vor}). This finding is in agreement with the recent results of the three-fluid hydrodynamical model~\cite{IS17}. However, physical details of this calculation are a little bit different: the results in~\cite{IS17} are given for baryonic fluids whereas we treat the matter including also mesons and partons. This difference is not very essential at the moderate colliding energy under consideration.

Multiplicities of strange particles and antiparticles are presented in Fig.~\ref{Lambda-num}. The calculations are performed with and without the account for the chiral symmetry restorations (CSR), cf. Ref.~\cite{PCSSMB16}. Besides $\Sigma^*$ and $\overline{\Sigma^*}$ rapidly decaying into $\Lambda$ or $\overline{\Lambda}$, the strange and anti-strange hyperons increase smoothly with time and differ roughly by two orders of magnitude. Note that multiplicity of (anti)$\Lambda$ hyperons includes both direct and those coming from the resonance decay. As demonstrated in Ref.~\cite{PCSSMB16}, the inclusion of the CSR provides a microscopic explanation for the ``horn'' structure in the excitation function of the $K^+/\pi$ ratio: the CSR in the hadronic phase produces a steep increase of this particle ratio up to $\sqrt{s_{NN}}\approx $7 GeV, while the drop at higher energies is associated with the appearance of deconfined partonic medium~\cite{PCSSMB16}. At the colliding energy considered the PHSD model accounts for the prediction of a fast growth of $\Lambda$ and $\bar\Lambda$  with time and then their flattening at about 20 fm/c. Contrary, the multiplicity of $\Sigma$ hyperons and anti-hyperons smoothly decreases in time. The conventional PHSD model~\cite{PHSD} without the CSR effect (lower panels in Fig.~\ref{Lambda-num}) provides similar qualitative behavior but the absolute  multiplicity of $\Lambda$ and $\bar\Lambda$ is lower by a factor of about two.

\section{Polarization treatment and results}

The mean spin vector of a particle of mass $m$ and spin $s$, produced around the point $x$ with the four-momentum $p$, in the leading order of thermal vorticity~\cite{Bec13,Bec15} is
\begin{align}
S^\mu(x,p)=-\frac{s(s+1)}{6m} (1\pm n(x,p))\varepsilon^{\mu\nu\lambda\delta}\
\varpi_{\nu\lambda}\ p_\delta\,,
\label{S-def}
\end{align}
where $n(x,p)$ is the Bose/Fermi distribution function and the Levi-Civita symbol $\varepsilon^{\mu\nu\lambda\delta}$ satisfies $\varepsilon^{0123}=1$.

This result may be directly  applied to a primary $\Lambda$ particle and we obtain for the 4-vector $S_\Lambda^\mu$
\begin{align}
S_\Lambda^\mu &=(S_\Lambda^0,\vec{S}_\Lambda)
\nonumber\\
&=\frac{1- n_\Lambda}{8m_\Lambda}\big( \vec{p}_\Lambda\cdot  \rot\vec{\beta},
E_\Lambda \rot\vec{\beta}+ [\vec{p}_\Lambda\times\vec{\varpi}_0] \big)\,,
\label{S-def-exp}
\end{align}
where $m_\Lambda$, $p_\Lambda$ and $E_\Lambda=\sqrt{m_\Lambda^2+p_\Lambda^2}$
are the mass, momentum and energy of the $\Lambda$ particle.

The magnitude of spin polarization of $\Lambda$ particles is determined by the asymmetry of the momentum distribution of daughter protons produced in decays $\Lambda\to p+\pi^-$ which in the $\Lambda$ rest frame can be parameterized as
\begin{align}
4\pi \frac{\rmd N}{\rmd\Omega^*}=1+\alpha_\Lambda \vec{P}_\Lambda^*\vec{n}_p^* \,,
\end{align}
where $\vec{P}^*$ is the polarization vector related to the spin
vector as
\begin{align}
\label{Pol-v}
\vec{P}^*_\Lambda=2\vec{S}^*_\Lambda
\end{align}
and $\vec{n}_p^*$ is the unit vector in the proton momentum direction both calculated in the $\Lambda$ rest frame, and
$\alpha_\Lambda=-\alpha_{\bar{\Lambda}}=0.642$ is the $\Lambda$ non-leptonic
decay constant. Boosting the 4-vector $S^\mu_\Lambda$ to the $\Lambda$ rest frame, we obtain that the zeroth component vanishes identically, $S_0^*=0$, and the spatial component becomes
\begin{align}
\vec{S}^*_\Lambda&= \vec{S}_\Lambda+\vec{S}_\Lambda\cdot\vec{p}_\Lambda  \frac{\vec{p}}{m_\Lambda(E_\Lambda+m_\Lambda)}
-S_{\Lambda}^0\,\frac{\vec{p}}{m_\Lambda}
\nonumber\\
&=\vec{S}_\Lambda-\vec{S}_\Lambda\cdot\vec{p}_\Lambda  \frac{\vec{p}_\Lambda}{E_\Lambda(E_\Lambda+m_\Lambda)},
\end{align}
where in the last equation we used the relation $S^0_\Lambda E_\Lambda=\vec{S}_\Lambda\cdot\vec{p}$, obviously following from (\ref{S-def-exp}).
Using (\ref{S-def-exp}) we can write explicitly
\begin{align}
\vec{S}^*_\Lambda=&\frac{1 - n_\Lambda}{8m_\Lambda }\Big( E_\Lambda \rot\vec{\beta} \nonumber \\
&+ 2[\vec{p}_\Lambda\times\vec{\varpi}_0]  + \vec{p}_\Lambda\cdot \rot\vec{\beta} \frac{\vec{p}_\Lambda}{(E_\Lambda + m_\Lambda)}
\Big)\,.
\label{S*-full}
\end{align}

In sampling the experimental data, one sums over the direction of
$\vec{p}_\Lambda$. The vector $\vec{S}^*$ averaged over the
$\vec{p}_\Lambda$ direction takes a very simple form
\begin{align}
\langle \vec{S}^*_\Lambda\rangle_{\vec{n}_p} &=
 \frac{(1 - n_\Lambda)}{4M_\Lambda }\Big( E_\Lambda
+  \frac13\frac{\vec{p\,}_\Lambda^2}{E_\Lambda + m_\Lambda}
\Big)\, \rot\vec{\beta}\,.
\label{S*}
\end{align}

A sizable amount of the final $\Lambda$'s is a product of resonance decays. In decays, the $\Lambda$s inherit a fraction of polarization of the initial (parent) states.
The spin vector of the parent state can be calculated using expressions (\ref{S*-full}) or (\ref{S*}) with the replacement of the $\Lambda$ mass, momentum and energy replaced by the corresponding quantities of the parent hyperon state. Additionally, one has to take into account the spin degeneracy factor and multiply the expression by $\frac{4}{3} s_P(s_P+1)$, where $s_P$ is spin of the parent state. For example for the spin-3/2 hyperons $\Sigma^*$ and $\Xi^*$ it will give the factor 5.

The main sources of secondary $\Lambda$s in our case are electromagnetic decays
$\Sigma^0\to \Lambda +\gamma$, strong decays $\Sigma^*\to \Lambda+\pi$, and $\Xi\to\Lambda\pi$ and sequential processes
$\Sigma^*\to \Sigma+\pi \to\Lambda+\pi+\gamma$ and $\Xi^*\to \Xi+\pi
\to \Lambda+\pi+\pi$. Thus, the number of secondary $\Lambda$'s produced in the $\Sigma^*$, $\Sigma$, $\Xi$'s and $\Xi^*$ decays can be calculated as
\begin{align}
&N_{\Lambda}^{(\rm sec.)}= N_{\Lambda}^{(\Sigma)} + N_{\Lambda}^{(\Sigma^*)} +  N_\Lambda^{(\Xi)}+ N_\Lambda^{(\Xi^*)}\,,
\label{NL-sec}\\
&N_{\Lambda}^{(\Sigma^*)}=B_{\Lambda\Sigma^*}N_{\Sigma^*} + B_{\Sigma\Sigma^*}(N_{\Sigma^{*+}} + N_{\Sigma^{*-}})/2\,,
\nonumber\\
&N_\Lambda^{(\Xi)}=B_{\Lambda\Xi}\,N_{\Xi}\,,
\,\,
N_\Lambda^{(\Xi^*)}=B_{\Lambda\Xi}N_{\Xi^{*}} \,, \,\,
N_{\Lambda}^{(\Sigma)}=N_{\Sigma^0}\,,
\nonumber
\end{align}
where $N_{\Sigma^*}=N_{\Sigma^{*+}}+ N_{\Sigma^{*0}} + N_{\Sigma^{*-}}$\,,
$N_\Xi=N_{\Xi^0} + N_{\Xi^-}$\,,
$N_{\Xi^*}=N_{\Xi^{*0}} + N_{\Xi^{*-}}$\,,
and $B_{H_fH_i}$ is the branching ratios for the transitions
$H_i\to H_f+\dots$ between the initial ($H_i$) and final ($H_f$)
hyperons. In Eq.~(\ref{NL-sec}) we take into account that the branching ratios $B_{\Lambda\Sigma^0}$ and $B_{\Xi\Xi^*}$ are equal to one and that $\Sigma^{*0}$ does not decay in $\Sigma^0+\pi^0$. For other branching ratios we have from~\cite{PDG} $B_{\Lambda\Sigma^*}=0.870$\,, $B_{\Sigma\Sigma^*}=0.117$\,, and $B_{\Lambda\Xi}=0.995$\,.
Relations similar to Eq.~(\ref{NL-sec}) hold also for anti-hyperons.

As argued in~\cite{Bec15}, the polarization of a daughter ($D$) baryon is proportional to the polarization  of a parent ($P$) baryon $\vec{S}^*_D = C_{DP}\vec{S}^*_P$, where $C_{DP}$ is a spin recoupling coefficient. For strong
and electromagnetic decays $C_{DP}$ is found in~\cite{KBec17,Bec15}
to be independent of the decay kinematics with the result $C_{\Xi\Xi^*}=C_{\Lambda\Sigma^*}=C_{\Sigma\Sigma^*}=\frac13$ and $C_{\Lambda\Sigma^0}=-\frac13$\,, whereas in weak decays of $\Xi$
the recoupling coefficient does depend on the decay kinematics and $C_{\Lambda\Xi^-}=0.927$\,, and $C_{\Lambda\Xi^0}=0.900$\,.
Thus, the averaged polarization of secondary $\Lambda$ particles can be calculated as
\begin{align}
\vec{S}_\Lambda^{\rm (sec.)} &=  \vec{S}_\Lambda^{\rm (\Sigma)}+ \vec{S}_\Lambda^{\rm (\Sigma^*)} + \vec{S}_\Lambda^{\rm (\Xi)} + \vec{S}_\Lambda^{\rm (\Xi^*)}\,,
 \label{SL-sec}\\
\vec{S}_\Lambda^{\rm (\Sigma)} &= C_{\Lambda\Sigma^0}p_{\Sigma^0}\vec{S}_\Sigma\,
\nonumber\\
\vec{S}_\Lambda^{\rm(\Sigma^*)} &= \big[C_{\Lambda\Sigma^*} B_{\Lambda\Sigma^*} p_{\Sigma^{*}}
\nonumber\\
&+\frac12 C_{\Lambda\Sigma^0} C_{\Sigma\Sigma^*}B_{\Sigma\Sigma^*}(p_{\Sigma^{*+}} + p_{\Sigma^{*-}})\big] \vec{S}_{\Sigma^*},
\nonumber\\
\vec{S}_\Lambda^{\rm (\Xi)} &= B_{\Lambda\Xi} (C_{\Lambda\Xi^0}p_{\Xi^0} + C_{\Lambda\Xi^-} p_{\Xi^-}) \vec{S}_{\Xi}\,,
\nonumber\\
\vec{S}_\Lambda^{\rm (\Xi^*)} &= \frac13 B_{\Lambda\Xi} C_{\Xi\Xi^*}
\big[(C_{\Lambda\Xi^0}+2 C_{\Lambda\Xi^-})
p_{\Xi^{*0}}
\nonumber\\
&+ ( C_{\Lambda\Xi^-} + 2 C_{\Lambda\Xi^0})  p_{\Xi^{*-}}\big]  \vec{S}_{\Xi^*}\,.
\nonumber
\end{align}
where $p_H$ is a relative contribution of hyperon $H$ to the total number of $\Lambda$s, $p_H=N_H/(N_\Lambda+N_\Lambda^{\rm (sec.)})$. The averaged contribution of primary $\Lambda$s is then given by $\vec{S}_\Lambda^{\rm (prim.)} = \vec{S}_\Lambda\, p_\Lambda$.
The same relations are valid also for anti-hyperons.

\begin{figure}
\centering
\includegraphics[width=8cm,clip]{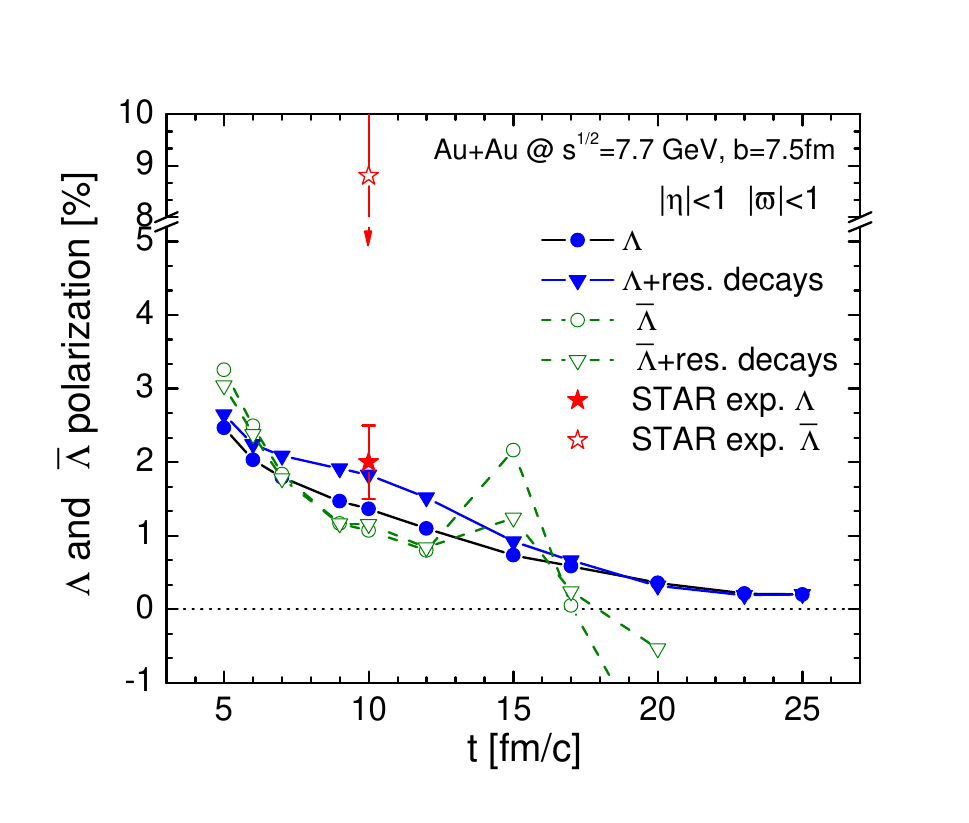}
\caption{(Color-online) Time dependence of the average hyperon polarization in peripheral Au+Au collisions. Full circles and triangles correspond to the primary and resonance decaying
$\Lambda$, respectively,  while similar but empty marks correspond
to $\bar\Lambda$. Stars with error bars are  experimental data for Au+Au collisions at $\sqrt{s_{NN}}=$7.7 GeV~\cite{BES}.  }
\label{Pol-v-t}
\end{figure}

Taking into account a possibility of  multi-step two-body decays,
we write the mean spin vector of primary+feed-down $\Lambda$s and the corresponding polarization as
\begin{align}
&\vec{S}_{\Lambda,\rm tot}^*=\frac12\vec{P}^*_{\Lambda,\rm tot}=\vec{S}_\Lambda^{\rm (prim.)} + \vec{S}_\Lambda^{\rm (sec.)}.
\label{SP-tot}
\end{align}
The average polarization vector calculated within the PHSD model, (\ref{Pol-v}) for Au+Au($\sqrt{s}=7.7$\,GeV) collisions and centrality $20-50\% $ is plotted in Fig.~\ref{Pol-v-t} for different moments of time.
For the considered reaction we compute the global $\Lambda$ polarization and estimate the $\Lambda's$ feed-down from resonance decays, cf. Eq.~(\ref{SP-tot}). The experimental cut $|\eta_\Lambda|\le|$ 1 is taken into account at a fixed time moment in such a way that it does not influence on the subsequent hadron evolution. At time  $t\approx$10 fm/c, the projection of the $\Lambda$ polarization onto the direction of the global angular momentum in off-central collisions, $P_{\Lambda,\rm tot}^*\approx 2\%$ which is  nicely close to the experimental value $2.\pm 0.6\%$, cf. Ref.~\cite{BES}, with the feed-down factor about 25\%. As to $\overline{\Lambda}$, none of the available models can predict correctly $P^*_{\overline{\Lambda},\rm tot}$ which is close or even higher than $P_{\Lambda,\rm tot}$. The energy $\sqrt{s}=7.7$\,GeV is of particular interest. Here the measured $P^*_{\overline{\Lambda},\rm tot}=8.7\pm 3.5$\% is four times larger than $P^*_{\Lambda,\rm tot}$~\cite{BES} and fluctuates at later time of interaction.

\section{Conclusions}

An analysis of vorticity within the kinetic PHSD model was performed for peripheral Au+Au collisions at the energy  $\sqrt{s_{NN}}=$7.7
GeV. The relativistic vorticity reaches a maximum soon after local equilibrium when the rotation equilibrates in the system. Then, similarly to other model considerations, the vorticity decreases rapidly due to explosive expansion of the system, still at $\approx 5\, {\rm fm}/c$ after the beginning of fluid dynamical expansion. Transition to the analysis in terms of the thermal vorticity gives larger values even at ultrarelativistic RHIC and LHC energies.
A similar study was performed recently~\cite{BGST13,TU15} in the QGSM approach. In the PHSD model the vorticity is oriented  in the $-y$ direction and
the result is maximal transverse polarization for particles emitted
in the reaction plane in the $(+/-)x$ direction while the polarization of particles emitted into the perpendicular $(+/-)y$ direction is negligible. In the case of chiral vortaic effect  with time significant helicity enhancement is expected for particles emitted in the $(+/-)y$ direction.

The calculated global polarization of $\Lambda$ in midrapidity region is  close to  the measured one but $\bar\Lambda$ polarization is strongly underestimated.

We plan to extend this kind of calculations to higher energies and different centralities in order to determine the best conditions for vorticity formation in relativistic nuclear collisions.

\section*{Acknowledgements}
We are thankful to M.~Baznat, F.~Becattini, E.~Bratkovskaya, Yu.~Ivanov, I.~Karpenko, A.~Khvorostukhin, O.~Rogachevsky, O.~Teryaev, and  G.~Zinovjev for helpful discussions and valuable comments.
The work is supported by Slovak Grant No.~1/0348/18 and by
THOR COST Action~CA15213. The authors acknowledge the support by
grant of the Plenipotentiary of the Slovak Government to JINR.

\end{document}